\documentclass{skaox2006}
\usepackage{graphicx}
\begin{document}
   \title{Future \ion{H}{i} Surveys on the road to the SKA }

   \author{Robert Braun\inst{1}
          }

   \institute{ASTRON, PO Box 2, 7990 AA Dwingeloo, The Netherlands}

   \abstract{In this short contribution we consider what types of
surveys might be optimally pursued with path-finding instruments of
1\%, 10\% and finally 100\% of the projected SKA sensitivity from the
perspective of scientific applications that utilize the red-shifted
$\lambda$21~cm emission line. Achieving interesting \ion{H}{i} galaxy
sample sizes with 1\% SKA surveys requires very substantial survey
durations, of about 1000 days. Good sampling (log(N)$\sim$5 down to
below M$_{HI}^*$) can then be achieved out to z~=~0.2 over 8000
deg$^2$ of survey area or even to z~=~0.5 over 800 deg$^2$. The same
surveys will permit the resolved imaging of order 1000 galaxies in
each of several red-shift bins as well as detection of faint neutral
filaments in the vicinity of galaxies with a column density of about
10$^{18}$cm$^{-2}$. Once 10\% SKA sensitivities are achieved, then
ground-breaking surveys are possible with only 100 day
duration. Sample sizes of log(N)$\sim$6 extending below M$_{HI}^*$ are
possible over 800 deg$^2$ out to z~=~0.5 and over 80 deg$^2$ out to
z~=~1. Such surveys will permit very competitive measurement of
acoustic oscillations in the galaxy power spectrum. One can then
envision a series of 10\% SKA surveys probing different depths. The
diffuse \ion{H}{i} sensitivity would be such
($\sim$10$^{17}$cm$^{-2}$) that the next factor of three in sky area
will become accessible to imaged detection and kinematic study of the
\ion{H}{i} cosmic web. With the 100\% SKA sensitivity the capabilities
are truly phenomenal. Survey sample sizes in the range log(N)~=~7--8
are feasible over the red-shift range of 0.2 to about 5. Precise
tracking of potential time evolution of dark energy (via the baryonic
acoustic oscillation signature) should be possible out to
z~$\sim$~3. The local cosmic web will be imaged down to
N$_{HI}$~=~10$^{16}$cm$^{-2}$. What exactly will be seen at z~$>$~3?
This will depend crucially on the SKA sensitivity in the critical
frequency window of 350 to 200~MHz. }

\maketitle
%
%
\section{Introduction}

One of the original (and still most compelling) motivations for
building the SKA was realizing the ability to study the evolving
neutral gas content of galaxies throughout cosmic time. The
distribution and kinematics of \ion{H}{i} as traced by the red-shifted
$\lambda$21~cm emission line provide unique insights into galaxy
formation and evolution. In addition to permitting direct assessment
of the atomic and dynamical mass, \ion{H}{i} imagery retains the
signature of galaxy interactions from the previous Gyr, rather than
only providing a snap-shot of gas content. However, current
instrumentation permits \ion{H}{i} detection and imaging in only the
very local universe. It will require some two orders of magnitude
greater instantaneous sensitivity to push back the \ion{H}{i} frontier
to the early universe.

Realization of the complete SKA is envisioned to take place by about
2020. However, a series of path-finding instruments will become
available as early as 2009, and these should pave the way for a 10\% SKA by
about 2015. What exactly might these instruments make possible and
what types of surveys might be optimally pursued with each? In this
short contribution we will consider this question from the perspective
of \ion{H}{i} surveys.

\section{Surveys}

The complete SKA is currently specified as providing 20000~m$^2$/K of
effective sensitivity (between at least 0.5 and 5 GHz) with an
instantaneous FOV significantly exceeding 1~deg$^2$ at low
frequencies. About 25\% of the collecting area is envisaged to be
within a region of 1~km diamter, 50\% within 5~km, and 75\% within
150~km. For simplicity, we will assume that the effective sensitivity
of the complete SKA for \ion{H}{i} galaxy surveys will be about
10000~m$^2$/K at all frequencies, while applications requiring the
highest brightness sensitivity will have 5000~m$^2$/K. This should be
accurate out to at least z~=~3, since at lower frequencies it will be
possible to use longer baselines without over-resolving targets, but
will certainly no longer apply at z~=~6, without a higher sensitivity
than currently envisaged between 200 and 300~MHz.

The currently funded path-finder instruments: APERTIF (APERture Tile
In Focus) in the Netherlands, xNTD (the eXtended New Technology
Demonstrator) in Australia and KAT (Karoo Array Telescope) in South
Africa each amount to about a 1\% SKA in terms of their surveying
power. APERTIF, by placing Focal Plane Array (FPA) receivers in each
of the 14 dishes of 25~m diameter that make up the WSRT array will
provide 100~m$^2$/K sensitivity over a 8~deg$^2$ FOV; while each of
xNTD and KAT is planned to provide about 50~m$^2$/K sensitivity over a
22~deg$^2$ FOV. Since survey speed scales as
BW$\times$FOV$\times$Sens$^2$, all three of these systems will provide
similar survey performance, given a similar instantaneous bandwidth of
about 300~MHz and frequency tuning range of at least 850 to 1700 MHz.

The distinguishing features of these SKA path-finders are: (1)
simultaneous, wide-field, wide-band data acquisition, followed by (2)
parallel multi-topical astrophysical analysis. Specialized science
teams (continuum, polarimetry, spectral line, transients, etc.) will
capitalize on the full survey potential and maximize the scientific
return on these instrument investments. This should permit world-class
science to be carried out in the decade preceding full SKA deployment.

Some of the major scientific themes that drive \ion{H}{i} science to higher
sensitivities are:
\begin{enumerate}
\item Quantifying the evolving gas content as well as the baryonic power
  spectrum of galaxies by detecting statistical samples spanning the
  widest possible range of red-shifts.

\item Witnessing galaxy formation and evolution via resolved imaging
  studies at the highest possible look-back times.

\item Imaging the local cosmic web by pushing back the N$_{HI}$
  frontier into the optically thin (to ionizing photons) regime below
  10$^{18}$cm$^{-2}$. 

\end{enumerate}

We have assessed the types of \ion{H}{i} surveys which might be
carried out to address these three science themes by instruments
having 1, 10 and 100\% of the full SKA sensitivity (taken to be
10000~m$^2$/K for galaxy surveys and 5000~m$^2$/K for the low N$_{HI}$
application) in order to determine their relative and absolute
utility. We have assumed an unevolved HIPASS HIMF (Zwaan et
al. \cite{zwaa03}) at all red-shifts to allow a conservative
prediction of galaxy detection rates. We further assume that the
effective line-width of the \ion{H}{i} signal is given by $\Delta
V(M_{HI})~=~0.105 M_{HI}^{1/3}$ in statistical agreement with local
galaxy populations. The assumed comological parameters are H$_0$~=~73
km/s/Mpc, $\Omega_m$~=~0.24 and $\Omega_\Lambda$~=~0.76. We also
assume that the facility has a instantaneous FOV of 8~deg$^2$
independent of frequency (as will be the case for APERTIF) and then
consider surveys which cover a total area on the sky of 8000, 800, 80
and 8~deg$^2$. The total observing time for each survey was 1000~days
for the 1\% SKA cases and 100~days for 10 and 100\% SKA cases. As will
be seen below, these appear to be realistic survey durations to reach
interesting depths.

The galaxy results are summarized in Tables~\ref{tab:dete} and
\ref{tab:imag}, where we have assumed a 7$\sigma$ threshold (at a
velocity resolution matched to the line-width) for simple detection
and a 100$\sigma$ threshold for imaging. This second criterion stems
from allowing for some 10's of high significance resolution elements
across each source.  We tabulate the (base 10) logarithm of the number
of detections (rounded to the nearest integer) of each survey in a
sequence of red-shift bins. 

\begin{figure*}
   \centering
   \includegraphics[width=14.8cm]{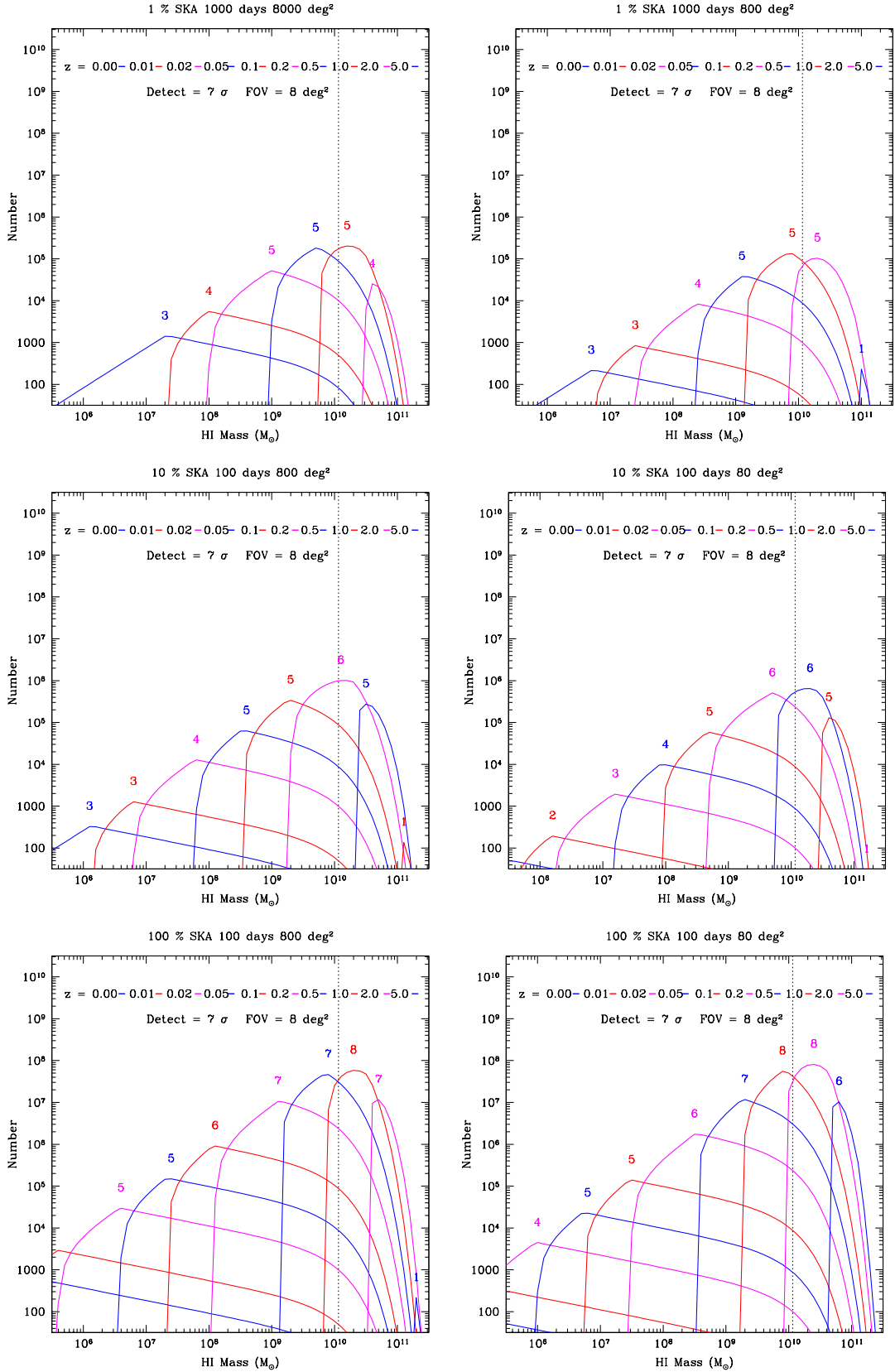}
      \caption{Detected numbers of galaxies as a function of their
      \ion{H}{i} mass for representative surveys. The different curves
      correspond to the red-shift intervals indicated at the top of the
      plots. The logarithm of the integrated number of detections in
      a red-shift interval is indicated by the integer above the
      peak of each curve. A detection threshold of 7$\sigma$ is
      assumed and an instantaneous FOV of 8~deg$^2$. The dotted
      vertical line is M$_{HI}^*$. The label above each panel
      gives the survey sensitivity, duration and total area. }
         \label{fig:surv}
   \end{figure*}

Several of the detection surveys are illustrated in more detail in
Fig.~\ref{fig:surv}. The number of detections per half dex \ion{H}{i}
mass bin are plotted in the figure. Separate curves are drawn for each
of the red-shift intervals to permit assessment of the achieved mass
depth in each interval. The dotted vertical line near
M$_{HI}$~=~10$^{10}$~M$_{\odot}$ is M$_{HI}^*$ of the HIMF. Good
sampling of a red-shift interval demands a significant detection rate
down to below M$_{HI}^*$.

The results for low column density surveys are summarized in
Table~\ref{tab:cweb} where the 1$\sigma$ column density sensitivity is
indicated over a 20 km~s$^{-1}$ line-width for a beam size of
60~arcsec. This beam size correponds roughly to that of the central km
of an array configuration. Such a beam size might reasonably be
expected to be filled with diffuse \ion{H}{i} emission out to
distances of about 30~Mpc where it subtends less than 10~kpc. At
larger distances, the effective column density sensitivity will likely
be diminished due to beam dilution.
 
\begin{table*}
\caption{Survey Size and Detection Results}             
\label{tab:dete}      
\centering          
\begin{tabular} {c c c c c c c c c c c c c }     
\hline\hline       
Sensitivity & Time & Area &\multicolumn{10}{c} {log(Number Detections)
  in z-range $>7\sigma$}\\
(\% SKA)& (Days) & (deg$^2$) & 0--0.01&0.01--0.02&0.02-0.05&0.05--0.1&0.1--0.2&0.2--0.5&0.5--1&1--2&2--5&$>$5\\ 
\hline           
1 & 1000 & 8000 & 3&4&5&5&5&4&-&-&-&-\\
1 & 1000 & 800 & 3&3&4&5&5&5&1&-&-&-\\
1 & 1000 & 80 & 2&2&3&4&4&5&4&-&-&-\\
1 & 1000 & 8 & 1&1&2&3&4&5&5&4&-&-\\
\hline           
10 & 100 & 8000 & 4&4&5&6&6&6&2&-&-&-\\
10 & 100 & 800 & 3&3&4&5&5&6&5&1&-&-\\
10 & 100 & 80 & 2&2&3&4&5&6&6&5&1&-\\
10 & 100 & 8 & 1&2&3&3&4&5&5&6&5&-\\
\hline           
100 & 100 & 8000 & 4&4&5&6&7&8&8&7&3&-\\
100 & 100 & 800 & 3&4&5&5&6&7&7&8&7&1\\
100 & 100 & 80 & 2&3&4&5&5&6&7&8&8&6\\
100 & 100 & 8 & 1&2&3&4&5&6&6&7&8&8\\
\hline                  
\end{tabular}
\end{table*}

\begin{table*}
\caption{Survey Size and Imaging Results }             
\label{tab:imag}      
\centering          
\begin{tabular} {c c c c c c c c c c c c c }     
\hline\hline       
Sensitivity & Time & Area &\multicolumn{10}{c} {log(Number Detections)
  in z-range $>100\sigma$}\\
(\% SKA)& (Days) & (deg$^2$) & 0--0.01&0.01--0.02&0.02-0.05&0.05--0.1&0.1--0.2&0.2--0.5&0.5--1&1--2&2--5&$>$5\\ 
\hline           
1 & 1000 & 8000 & 3&3&3&2&-&-&-&-&-&-\\
1 & 1000 & 800 & 2&2&3&3&2&-&-&-&-&-\\
1 & 1000 & 80 & 1&2&3&3&3&1&-&-&-&-\\
1 & 1000 & 8 & 1&1&2&2&3&2&-&-&-&-\\
\hline           
10 & 100 & 8000 & 3&3&4&4&3&-&-&-&-&-\\
10 & 100 & 800 & 2&3&4&4&4&2&-&-&-&-\\
10 & 100 & 80 & 2&2&3&3&4&3&-&-&-&-\\
10 & 100 & 8 & 1&1&2&3&3&4&2&-&-&-\\
\hline           
100 & 100 & 8000 & 4&4&5&5&6&5&-&-&-&-\\
100 & 100 & 800 & 3&3&4&5&5&6&4&-&-&-\\
100 & 100 & 80 & 2&2&3&4&5&5&5&4&-&-\\
100 & 100 & 8 & 1&2&3&3&4&5&5&5&3&-\\
\hline                  
\end{tabular}
\end{table*}

\begin{table}
\caption{Survey Size and Cosmic Web Depth }             
\label{tab:cweb}      
\centering          
\begin{tabular} {c c c c c }     
\hline\hline       
Sensitivity & Time & Area & log($\Delta N_{HI}$)\\
(\% SKA)& (Days) & (deg$^2$) & ($\theta$=60'', $\Delta$V=20 km/s)\\ 
\hline           
1 & 1000 & 8000 & 18.5\\
1 & 1000 & 800 & 18\\
1 & 1000 & 80 & 17.5\\
1 & 1000 & 8 & 17\\
\hline           
10 & 100 & 8000 & 18\\
10 & 100 & 800 & 17.5\\
10 & 100 & 80 & 17\\
10 & 100 & 8 & 16.5\\
\hline           
100 & 100 & 8000 & 17\\
100 & 100 & 800 & 16.5\\
100 & 100 & 80 & 16\\
100 & 100 & 8 & 15.5\\
\hline                  
\end{tabular}
\end{table}

%

\section{Conclusions}

Achieving interesting \ion{H}{i} galaxy sample sizes with 1\% SKA
surveys requires very substantial survey durations, of about 1000
days. Good sampling (log(N)$\sim$5 down to below M$_{HI}^*$) can then
be achieved out to z~=~0.2 over 8000 deg$^2$ or even to z~=~0.5 over
800 deg$^2$ as shown in Table~\ref{tab:dete}. The same surveys would
permit the resolved imaging of order 1000 galaxies in each of several
red-shift bins as can be seen from Table~\ref{tab:imag} as well as
detection of faint neutral filaments in the vicinity of galaxies with
a logarithmic column density {\sc RMS} of 18.5 or 18 (from
Table~\ref{tab:cweb}). This should be compared with the current
detection threshold in all but the deepest \ion{H}{i} imaging surveys
of about 10$^{19}$cm$^{-2}$ over $\Delta$V~=~20~km~s$^{-1}$. Since the
surface area of diffuse \ion{H}{i} is known to increase by a factor of
three for each decade of column density in this regime (from the
statistics of QSO absorption lines, cf. Braun \& Thilker
\cite{brau04}) it is clear that the first step towards systematic
mapping of the diffuse \ion{H}{i} filament distribution will be
achieved in such surveys. Simultaneous with the various \ion{H}{i}, OH
and other spectral line applications that would be served by these
surveys are the continuum, polarimetric and variability applications
that other members of a survey science team would exploit.

The very long survey durations, underline the great utility of having
several of such 1\% SKA path-finders operational in the same
timeframe; APERTIF in the North and xNTD and KAT in the
South. Complimentary surveys could then be carried out by the different
facilities to maximize the total scientific return.

Once 10\% SKA sensitivities are achieved, then ground-breaking surveys
are possible with only 100 day duration. Sample sizes of log(N)$\sim$6
extending below M$_{HI}^*$ are possible over 800 deg$^2$ out to
z~=~0.5 and over 80 deg$^2$ out to z~=~1. Such surveys will permit
very competitive measurement of acoustic oscillations in the galaxy
power spectrum (eg. Blake \& Glazebrook \cite{blak03}). Given the more
modest survey duration (relative to the 1\% SKA case), one can
envision a series of surveys probing different depths. The diffuse
\ion{H}{i} sensitivity is such ($\sim$10$^{17}$cm$^{-2}$) that the next
factor of three in sky area will become accessible to imaged detection
and kinematic study of the cosmic web.

With the 100\% SKA sensitivity the capabilities are truly
phenomenal. Survey sample sizes in the range log(N)~=~7--8 are
feasible over the red-shift range of 0.2 to about 5. The SKA will
easily be the most productive red-shift engine in astronomy. This
finally brings \ion{H}{i} imaging into completely new terrain. Precise
tracking of potential time evolution of dark energy (via the baryonic
acoustic oscillation signature) should be possible out to
z~$\sim$~3. The local cosmic web will be imaged down to
N$_{HI}$~=~10$^{16}$cm$^{-2}$. What exactly will be seen at z~$>$3?
This will depend crucially on the SKA sensitivity in the critical
frequency window of 350 to 200~MHz. If this can be maintained at the
level of 10000~m$^2$/K, then the prospects are extremely good for
detecting large populations of early universe objects. Given the very
low cost of collecting area in this frequency range, this seems to be
a very worthwhile area for investment.

\end{document}